\documentclass[aps,prl,superscriptaddress,twocolumn]{revtex4}
\usepackage{graphics}
\usepackage{amsmath}
\usepackage{graphicx}
\usepackage{amsfonts}
\usepackage{amssymb}

\begin{document}

\title{Continuous Variable Quantum Cryptography \\ using Two-Way Quantum Communication}
\author{Stefano Pirandola}
\affiliation{M.I.T. - Research Laboratory of Electronics,
Cambridge MA 02139, USA}
\author{Stefano Mancini}
\affiliation{Dipartimento di Fisica, Universit\`{a} di Camerino,
Camerino 62032, Italy}
\author{Seth Lloyd}
\affiliation{M.I.T. - Research Laboratory of Electronics,
Cambridge MA 02139, USA} \affiliation{M.I.T. - Department of
Mechanical Engineering, Cambridge MA 02139, USA}
\author{Samuel L. Braunstein}
\date{\today }
\affiliation{Computer Science, University of York, York YO10 5DD,
United Kingdom} \maketitle

\textbf{Quantum cryptography has been recently extended to
continuous variable systems, e.g., the bosonic modes of the
electromagnetic field. In particular, several cryptographic
protocols have been proposed and experimentally implemented using
bosonic modes with Gaussian statistics. Such protocols have shown
the possibility of reaching very high secret key rates, even in
the presence of strong losses in the quantum communication
channel. Despite this robustness to loss, their security can be
affected by more general attacks where extra Gaussian noise is
introduced by the eavesdropper. In this general scenario we show a
\textquotedblleft hardware solution\textquotedblright\ for
enhancing the security thresholds of these protocols. This is
possible by extending them to a two-way quantum communication
where subsequent uses of the quantum channel are suitably
combined. In the resulting two-way schemes, one of the honest
parties assists the secret encoding of the other with the chance
of a non-trivial superadditive enhancement of the security
thresholds. Such results enable
the extension of quantum cryptography to more complex quantum communications.%
}

In recent years, quantum information has entered the domain of
continuous variable (CV) systems, i.e., quantum systems described
by an infinite dimensional Hilbert space \cite{CVbook,BraReview}.
So far, the most studied CV systems are the bosonic modes, such as
the optical modes of the electromagnetic field. In particular, the
most important bosonic states are the ones with Gaussian
statistics, thanks to their experimental accessibility and the
relative simplicity of their mathematical description
\cite{EisertGauss,GaussianStates}. Accordingly, quantum key
distribution
(QKD) has been extended to this new framework \cite%
{Hillery,Ralph1,Ralph2,Reid,Preskill,Cerf0,Homo,Homo2,Cerf,Cerf2,Hetero,Hetero2,Silber,Hirano1,Hirano2,Hirano3,Lutk}
and Gaussian cryptographic protocols using coherent states have been shown
to exploit fully the potentialities of quantum optics \cite{Homo2,Hetero2}.
These coherent-state protocols are robust with respect to the noise of the
quantum channel, as long as such noise can be ascribed to pure losses \cite%
{Homo2,Hetero2}. By contrast, their security is strongly affected when
channel losses are used to introduce a thermal environment, which is assumed
to be controlled by a malicious eavesdropper \cite{Homo2,Estimators}. In
this Gaussian eavesdropping scenario, we present a method to enhance the
security thresholds of the basic coherent-state protocols. This is achieved
by extending them to two-way quantum communication protocols, where one of
the honest parties (Bob) uses its quantum resources to assist the secret
encoding of the other party (Alice). In particular, the enhancement of
security is proven to be effective since the security thresholds are
superadditive with respect to the double use of the quantum channel. Such a
result is achieved when the Gaussian attack corresponds to a memoryless
Gaussian channel. More generally, we also consider Gaussian channels with
memory, therefore creating classical and/or quantum correlations between the
paths of the two-way quantum communication. In order to overcome this kind
of eavesdropping strategy, the two-way protocols must be modified into
suitable hybrid protocols, which represent their safe formulation against
every kind of collective Gaussian attack.

\section{One-way protocols}

In basic coherent-state protocols \cite{Homo,Hetero}, Alice prepares a
coherent state $\left\vert \alpha \right\rangle $ whose amplitude $\alpha
=(Q_{A}+iP_{A})/2$ is stochastically modulated by a pair of independent
Gaussian variables $\{Q_{A},P_{A}\}$, with zero mean and variance $V-1$.
This variance determines the portion of phase space which is available to
Alice's classical encoding $\{Q_{A},P_{A}\}$ and, therefore, quantifies the
amount of energy which Alice can use in the process. This energy is usually
assumed to be very large $V\gg 1$ (large modulation limit) in order to reach
the optimal and asymptotic performances provided by the infinite dimensional
Hilbert space. The modulated coherent state is then sent to Bob through a
quantum channel, whose noise is assumed ascribable to the malicious action
of a potential eavesdropper (Eve). In a homodyne ($Hom$) protocol \cite{Homo}%
, Bob detects the state via a single quadrature measurement (i.e., by a
homodyne\textit{\ }detection). More exactly, Bob randomly measures the
quadrature $\hat{Q}$ \textit{or} $\hat{P}$, getting a real outcome $%
X_{B}=Q_{B}$ (or $P_{B}$) which is correlated to the encoded signal $%
X_{A}=Q_{A}$ (or $P_{A}$). In a heterodyne ($Het$) protocol \cite{Hetero},
Bob performs a joint measurement of $\hat{Q}$ \textit{and} $\hat{P}$ (i.e.,
a heterodyne detection). In such a case, Bob decodes the $\mathbb{R}^{2}$%
-variable $X_{B}=\{Q_{B},P_{B}\}$ correlated to the total signal $%
X_{A}=\{Q_{A},P_{A}\}$ encoded in the amplitude $\alpha $. In both cases,
Alice and Bob finally possess two correlated variables $X_{A}$ and $X_{B}$,
characterized by some mutual information $I(X_{A}:X_{B})$. In order to
access this mutual information, either Bob estimates Alice's encoding $X_{A}$%
\ via a \textit{direct }reconciliation (DR) or Alice estimates Bob's
outcomes $X_{B}$\ via a \textit{reverse }reconciliation (RR) \cite%
{Estimators}. However, in order to extract some shared \textit{secret}
information from $I(X_{A}:X_{B})$, the honest parties\ must estimate the
noise of the channel by broadcasting and comparing part of their data. In
this way, they are able to bound the information $I(X_{A}:E)$ or $I(X_{B}:E)$
which has been potentially stolen by Eve during the process. Then, the
accessible secret information is simply given by $R^{\blacktriangleright
}:=I(X_{A}:X_{B})-I(X_{A}:E)$ for DR and by $R^{\blacktriangleleft
}:=I(X_{A}:X_{B})-I(X_{B}:E)$ for RR. Such secret information can be put in
the form of a binary key by slicing the phase space and adopting the
standard techniques of error correction and privacy amplification \cite%
{GaussRec}. In particular, Alice and Bob can extract a secret key whenever
the channel noise is less than certain \emph{security thresholds}, which
correspond to the boundary conditions $R^{\blacktriangleright }=0$ and $%
R^{\blacktriangleleft }=0$.

\begin{figure}[tbph]
\vspace{-0.3cm}
\par
\begin{center}
\includegraphics[width=0.5\textwidth] {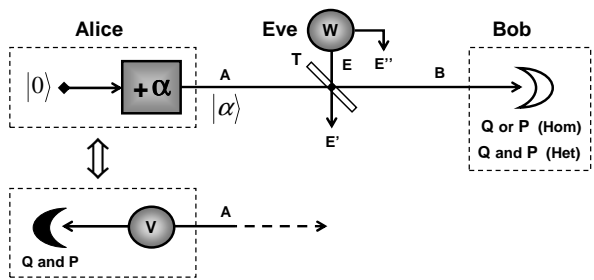}
\end{center}
\par
\vspace{-1.0cm}
\caption{Coherent state protocols. Alice prepares a coherent state $%
\left\vert \protect\alpha \right\rangle $ that Bob detects via a homodyne or
heterodyne detector. A Gaussian (entangling cloner) attack by Eve is also
shown. Note that preparing a coherent state (via modulating the vacuum) can
be equivalently achieved by heterodyning one of the two modes of an EPR\
pair.}
\label{OneWayPic}
\end{figure}

\begin{figure}[tbph]
\vspace{-0cm}
\par
\begin{center}
\includegraphics[width=0.44\textwidth] {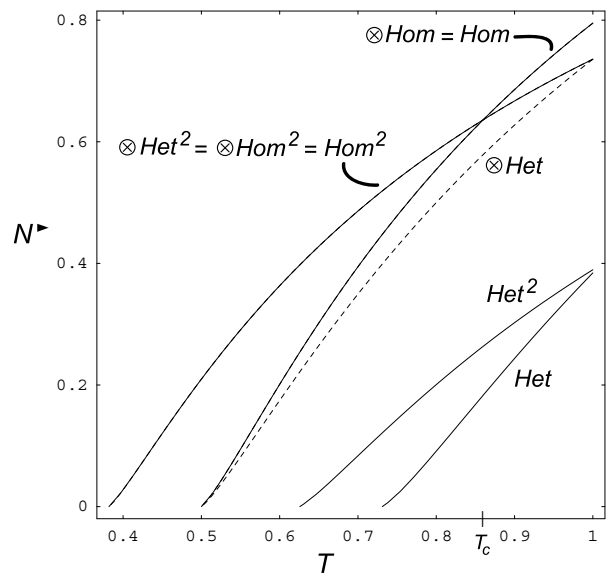}
\end{center}
\par
\vspace{-0.7cm}
\caption{Security thresholds in \textit{direct reconciliation}. The figure
shows the tolerable excess noise $N^{\blacktriangleright }$ (in quantum
shot-noise units) versus transmission $T$. Curves compare the various
one-way and two-way protocols against one-mode Gaussian attacks $\{N,T\}$ in
the limit of large modulation $(V\rightarrow +\infty )$.}
\label{DirectPic}
\end{figure}

In the CV framework, collective Gaussian attacks represent the
most powerful tool that today can be handled in the cryptoanalysis
of Gaussian state protocols \cite{AcinNew,CerfNew,Renner,Renner2}.
In the most general definition of a collective attack, all the
quantum systems used by Alice and Bob in a\emph{\ single run }of
the protocol are made to interact with a fresh ancillary system
prepared by Eve. Then, all the output ancillas, coming from a
large number of such single-run interactions, are subject to a
final coherent measurement that is furthermore optimized upon all
of Alice and Bob's classical communications. In particular, the
collective attack is Gaussian if the single-run interactions are
Gaussian, i.e., corresponding to unitaries that preserve the
Gaussian statistics of the states. Notice that for standard
one-way QKD, a single run of the protocol corresponds to a single
use of the channel. As a consequence, every collective Gaussian
attack against one-way protocols results in a memoryless channel
and, therefore, can be called a \emph{one-mode} Gaussian attack.
Since the quadratures encode independent variables
$\{Q_{A},P_{A}\}$, the single-run Gaussian interactions do not
need to mix the quadratures
\cite{AcinNew,CerfNew,GrangNew,RaulNew}. As a consequence, the
Gaussian interaction can be modelled by an entangling cloner
\cite{Homo2} (see Fig.~\ref{OneWayPic}) where a beam splitter (of
transmission $T$) mixes each signal mode $A$\ with an ancillary
mode $E$ belonging to an Einstein-Podolsky-Rosen (EPR) pair (see
Appendix). Such an EPR\ pair is characterized by a variance $W$
and correlates the two output ancillary modes $E^{\prime
},E^{\prime \prime }$ to be detected in the final coherent
measurement. Notice that, from the point of view of Alice and Bob,
this EPR pair simply
reduces to an environmental thermal state $\rho _{E}$ with thermal\ number $%
\bar{n}_{E}=(W-1)/2$. A one-mode Gaussian attack can be therefore described
by two parameters: transmission $T$ and variance $W$ or, equivalently, by $T$
and $N:=(W-1)(1-T)T^{-1}$, the latter being the \emph{excess noise} of the
channel. This parameter quantifies the amount of extra noise which is not
referable to losses, i.e., the effect of the thermal noise scaled by the
transmission \cite{Homo2}. The security thresholds against these powerful
attacks can be expressed in terms of\emph{\ tolerable} excess noise $%
\{N^{\blacktriangleright },N^{\blacktriangleleft }\}$ versus the
transmission $T$ of the channel. For protocols $Hom$ and $Het$, these
thresholds are displayed in Fig.~\ref{DirectPic} for DR and Fig.~\ref%
{ReversePic} for RR, and they confirm the results previously found in Refs.~%
\cite{LCcollective,LCcollective2} (see Appendix).

\begin{figure}[tbph]
\vspace{-0cm}
\par
\begin{center}
\includegraphics[width=0.42\textwidth] {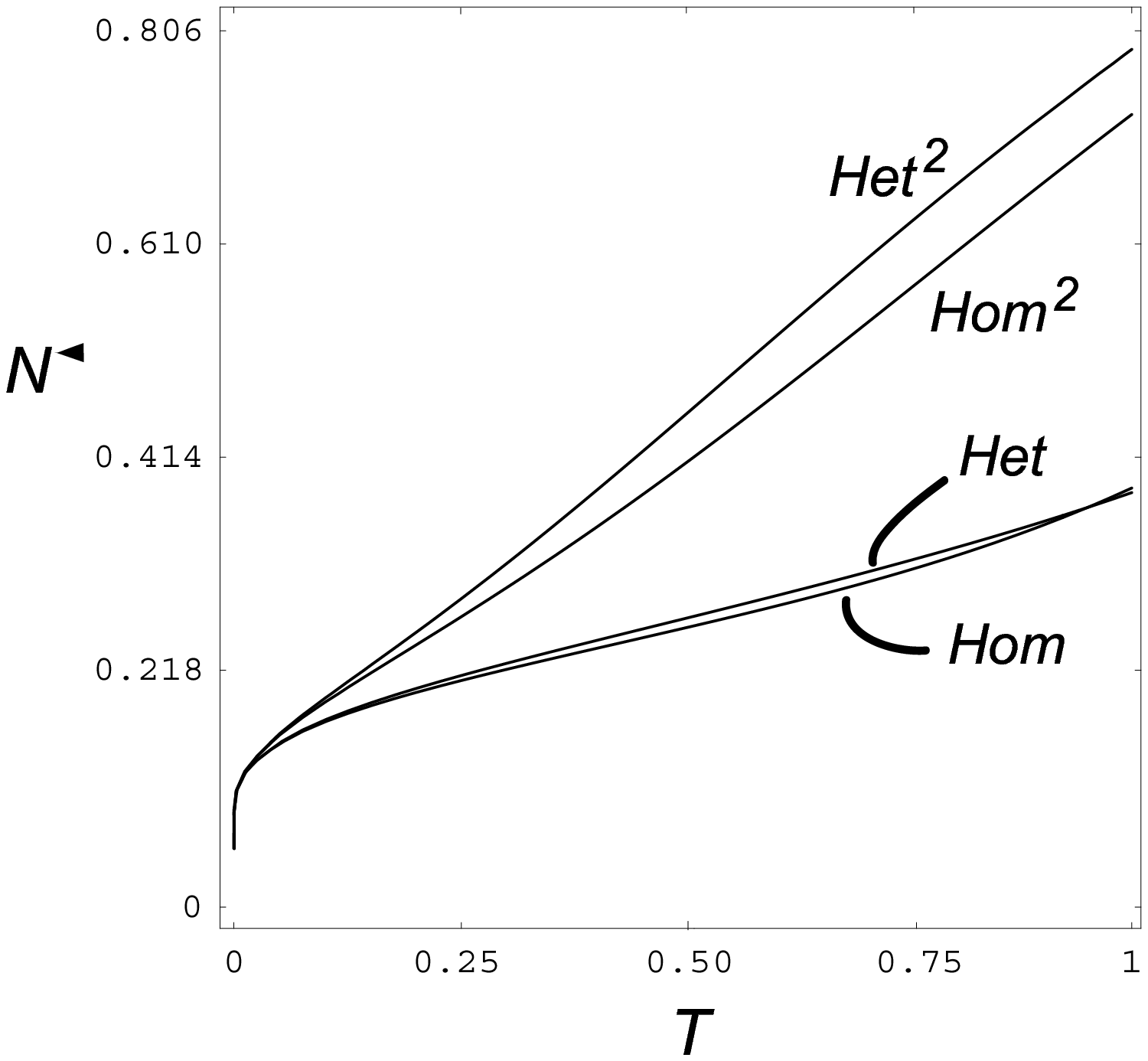}
\end{center}
\par
\vspace{-0.7cm}
\caption{Security thresholds in \textit{reverse reconciliation}. The figure
shows the tolerable excess noise $N^{\blacktriangleleft }$ (in quantum
shot-noise units) versus transmission $T$. Curves compare the individual
one-way and two-way protocols against one-mode Gaussian attacks $\{N,T\}$ in
the limit of large modulation $(V\rightarrow +\infty )$.}
\label{ReversePic}
\end{figure}

\section{From one-way to two-way protocols}

\begin{figure}[tbph]
\vspace{-0cm}
\par
\begin{center}
\includegraphics[width=0.47\textwidth] {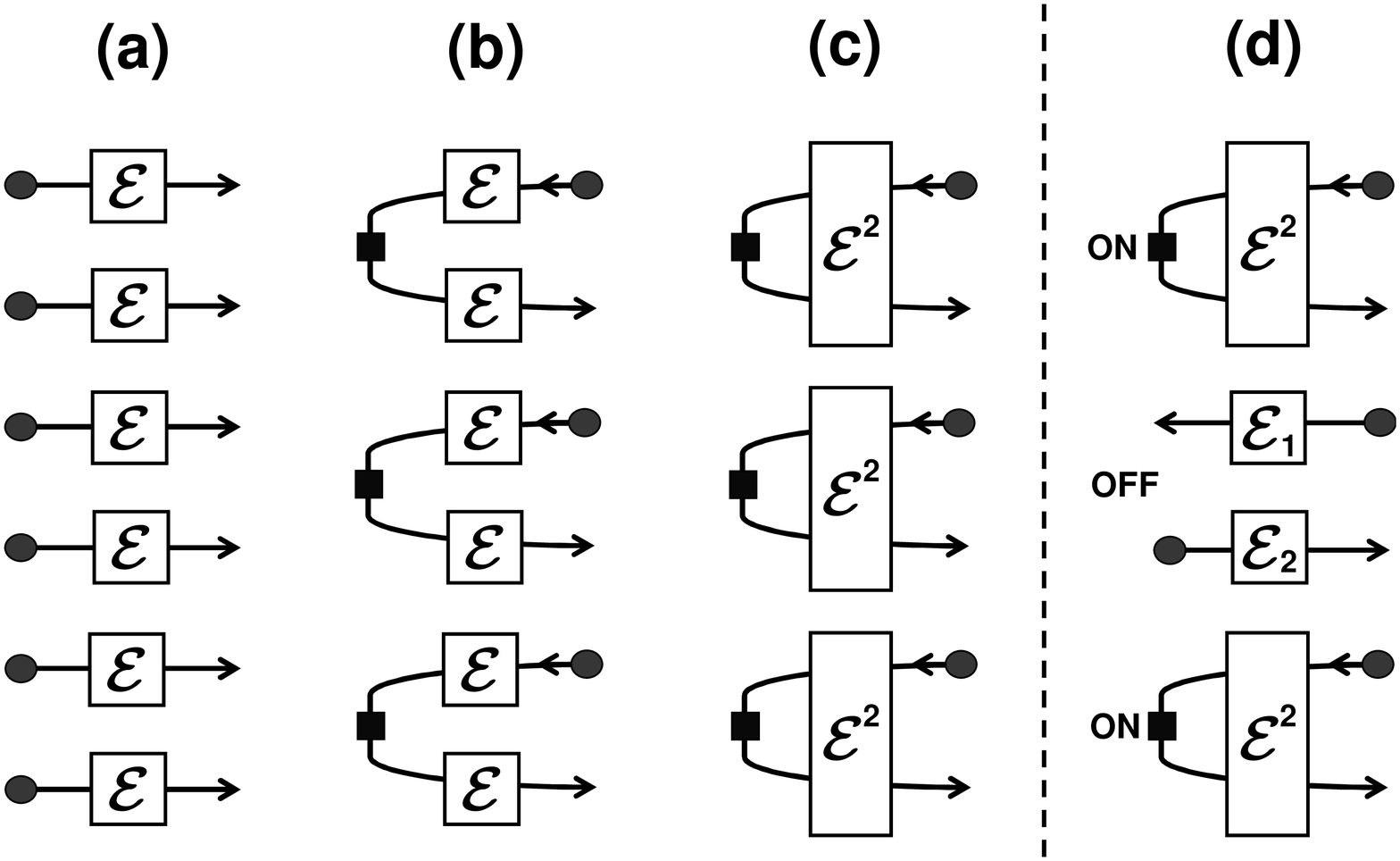}
\end{center}
\par
\vspace{-0.4cm}
\caption{General structure of the one-way, two-way and hybrid protocols,
together with their possible collective attacks. Circles and squares
represent the physical resources available in the process. In particular,
circles represent entanglement (possibly virtual) while squares are unitary
modulations. In the inset (a), the basic one-way scheme is depicted, where
all the resources are owned by Alice. In the insets (b) and (c), the
physical resources are instead distributed between Alice and Bob, where Bob
uses them for assisting while Alice for encoding (two-way scheme). Inset (d)
shows the hybrid protocol where one-way's (OFF) and two-way's (ON)\ are
randomly switched. All the insets also display Eve's collective attacks. The
inset (a) shows the collective attacks against one-way protocols (\emph{%
one-mode} attacks). The insets (b) and (c) show instead the collective
attacks against the two-way protocols. These are \emph{one-mode} (or \emph{%
reducible two-mode}) in the inset (b) while\emph{\ two-mode} in the inset
(c). Finally, inset (d) shows the effects of a two-mode attack on the hybrid
protocol.}
\label{Pic1}
\end{figure}

The above coherent state protocols have been simply formulated in terms of
prepare and measure (PM) schemes. Equivalently, they can be formulated as
entanglement-based schemes, where Alice and Bob extract a key from the
correlated outcomes of the measurements made upon two entangled modes (see
Fig.~\ref{OneWayPic}). In fact, heterodyning one of the two entangled modes
of an EPR pair (with variance $V$) is equivalent to remotely preparing a
coherent state $\left\vert \alpha \right\rangle $ whose amplitude is
randomly modulated by a Gaussian (with variance $V-1$) \cite{Estimators}. In
this dual representation of the protocol, Alice owns a physical resource
which can be equivalently seen as an amount of energy $\sim V$ for
modulation (in the PM representation) or as an amount of entanglement $\sim
\log 2V$ to be distributed (in the entanglement-based representation).
Because of this equivalence, the previous entanglement is also called \emph{%
virtual} \cite{Estimators}. In the above one-way protocols, all these
physical resources are the monopoly of Alice and their sole purpose is the
encoding of secret information. However, we can also consider a scenario
where these resources are symmetrically distributed between Alice and Bob,
and part of them is used to\emph{\ assist} the encoding. This is achieved by
combining Alice and Bob in a two-way quantum communication where Bob's
physical resources, to be generally intended as entanglement resources,
assist the secret encoding of Alice, which is realized by unitary random
modulations (see Fig.~\ref{Pic1}).

\begin{figure}[tbph]
\vspace{-0cm}
\par
\begin{center}
\includegraphics[width=0.5\textwidth] {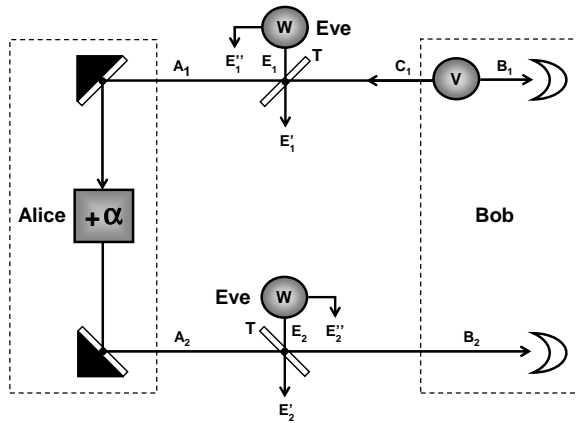}
\end{center}
\par
\vspace{-0.7cm}
\caption{Two-way quantum communication scheme. Bob exploits an EPR pair to
assist Alice's encoding. He keeps one mode $B_{1}$ while sending the other
one $C_{1}$ to Alice who, in turn, perfoms a stochastic phase-space
displacement $+\protect\alpha $. The resulting mode $A_{2}$ is then sent
back to Bob. The final modes $B_{1}$ and $B_{2}$\ are incoherently detected
by means of two homodyne detectors ($Hom^{2}$ protocol) or two heterodyne
detectors ($Het^{2}$ protocol). The figure also displays a one-mode Gaussian
attack, where the forward mode $C_{1}$ (from Bob to Alice) and the backward
one $A_{2}$ (from Alice to Bob)\ are subject to the action of entangling
cloners. }
\label{TwoWayPic}
\end{figure}

Let us explicitly construct such a two-way quantum communication. In simple
two-way generalizations, $Hom^{2}$ and $Het^{2}$, of the previous one-way
protocols, $Hom$ and $Het$, Bob exploits an assisting EPR pair (with
variance $V$) of which he keeps one mode $B_{1}$\ while sending the other to
Alice (see Fig.~\ref{TwoWayPic}). Then, Alice encodes her information via
Gaussian modulation (with variance $V-1$) by adding a stochastic amplitude $%
\alpha =(Q_{A}+iP_{A})/2$ to the received mode. Such a mode is then sent
back to Bob, where it is detected together with the unsent mode $B_{1}$.
Depending on the protocol, Bob will perform different detections on modes $%
B_{1}$ and $B_{2}$. In particular, for the $Hom^{2}$ protocol, Bob will
detect the $\hat{Q}$ (or $\hat{P}$) quadrature of such modes (homodyne
detections), while, for the $Het^{2}$ protocol, he will detect both $\hat{Q}$
and $\hat{P}$ (heterodyne detections). From the outcomes, Bob will finally
construct an optimal estimator $X_{B}$ of Alice's corresponding variable $%
X_{A}$, equal to $Q_{A}$ (or $P_{A}$) for $Hom^{2}$ and to the $\mathbb{R}%
^{2}$-vector $\{Q_{A},P_{A}\}$ for $Het^{2}$.

Since Bob's decoding strategy consists in individual \emph{incoherent}
detections, these entanglement-assisted QKD schemes are actually equivalent
to two-way schemes without entanglement, where Bob stochastically prepares a
quantum state to be sequentially transmitted forward and backward in the
channel. In fact, one can assume that Bob detects $B_{1}$ at the \emph{%
beginning} of the quantum communication, so that the travelling mode $C_{1}$
is randomly prepared in a \emph{reference}\ quantum state (which is squeezed
for $Hom^{2}$ and coherent for $Het^{2}$). This reference state reaches
Alice where stores the encoding transformation and, then, is finally
detected by the second decoding measurement of Bob. Therefore, if we
restrict Bob to incoherent detections ($\emph{classical}$ Bob), then the
two-way schemes also possess a dual representation, where the assisting
entanglement resource is actually virtual, i.e., can be replaced by an
equivalent random modulation. In this dual (entanglement-free)
representation, the advantage brought by the two-way quantum communication
can be understood in terms of an iterated use of the uncertainty principle,
where Eve is forced to produce a double perturbation of the same quantum
channel. For instance, let us consider the $Het^{2}$ protocol in the absence
of eavesdropping. By heterodyning mode $B_{1}$, Bob randomly prepares mode $%
C_{1}$ in a reference coherent state $\left\vert \beta \right\rangle $
containing a random modulation $\gamma $ known only to him. Then, Alice
transforms this state into another coherent state $\left\vert \alpha +\beta
\right\rangle $ which is sent back to Bob. By the subsequent heterodyne
detection, Bob is able to estimate the total amplitude $\alpha +\beta $ and,
therefore, to infer the signal $\alpha $ from his knowledge of $\beta $. If
we now insert Eve in this scenario, we see that she must estimate\emph{\ both%
} the reference $\beta $ and the masked signal $\alpha +\beta $ in order to
access the signal $\alpha $. This implies attacking both the forward and the
backward channel (see Fig.~\ref{TwoWayPic}) and, since the noise of the
first attack will perturb the second attack, we expect a non trivial
security improvement in the process. Such an effect intuitively holds under
the assumption of one-mode attacks (where the two paths are attacked
incoherently) and it is indeed confirmed by our analysis. Quantitatively, we
have tested the security performances of the two-way protocols against the
one-mode Gaussian attacks $\{N,T\}$ and the corresponding security
thresholds $N^{\blacktriangleright ,\blacktriangleleft
}=N^{\blacktriangleright ,\blacktriangleleft }(T)$ are shown in Fig.~\ref%
{DirectPic} and Fig.~\ref{ReversePic} (see Appendix). For the
two-way protocols, such thresholds relate the tolerable excess
noise to the transmission \textit{in each use of the channel} and,
therefore, they are directly comparable with the thresholds of the
corresponding one-way protocols. By comparing $Hom^{2}$ with $Hom$
and $Het^{2}$ with $Het$, one sees that the security thresholds
are improved almost everywhere (the only exception being $Hom^{2}$
for $T>T_{c}\simeq 0.86$ in DR). Such a superadditive behavior is
the central result of this work. Roughly speaking, even if two
communication lines (e.g., two optical fibers) are too noisy for
one-way QKD, they can be combined to enable a two-way QKD, as long
as the quantum channel is memoryless.

In order to deepen our analysis on superadditivity, we also tested
the previous one-way and two-way protocols when a classical Bob is
replaced by a \emph{quantum} Bob. This means that Bob is no longer
limited to incoherent detections but can access a quantum memory
storing all the modes involved in the quantum communication. Then,
Bob performs a final optimal coherent measurement on all these
modes in order to retrieve Alice's information. Such a coherent
measurement can be disjoint, i.e., designed to estimate a single
quadrature for each encoding, or joint, i.e., designed to estimate
both quadratures. Correspondingly, the modified one-way and
two-way protocols will be denoted by $\otimes Hom$, $\otimes Het$,
$\otimes Hom^{2}$ and $\otimes Het^{2}$. Notice that these
\emph{collective} protocols may not admit an equivalent
entanglement-free representation (where Bob's entanglement is
replaced by a random modulation) if Bob's coherent measurement
cannot be reduced to incoherent detection. The corresponding
security thresholds are shown in Fig.~\ref{DirectPic} (only DR can
be compared, see Appendix). It is evident that superadditivity
holds almost everywhere also for these collective schemes, the
only exception being $\otimes Hom^{2}$ above the same critical
value $T_{c}$\ as
before. We easily note that $\otimes Hom$ coincides with $Hom$, while $%
\otimes Hom^{2}$ coincides with $Hom^{2}$. Then, in the case of disjoint
decoding, the optimal coherent measurement asymptotically coincides with a
sequence of incoherent homodyne detections. As a consequence, the collective
protocols ($\otimes Hom$ and $\otimes Hom^{2}$) collapse to the
corresponding individual protocols ($Hom$ and $Hom^{2}$), where there is no
need for a quantum memory. In particular, this proves that $\otimes Hom^{2}$
admits an entanglement-free representation where infinitely-squeezed states
are sent to Alice through the forward path, and are then homodyned at the
output of the backward path. The usage of quantum memories does better in
the case of joint decoding, since $\otimes Het$ and $\otimes Het^{2}$ have
much better performances than the corresponding individual protocols $Het$
and $Het^{2}$. As a consequence, no simple entanglement-free representation
is known for $\otimes Het^{2}$.

\section{Hybrid protocols}

We remark that our previous quantitative cryptoanalysis concerns one-mode
Gaussian attacks, which are the cryptographic analog of a memoryless
Gaussian channel. However, when a multi-way scheme is considered, a single
run of the protocol no longer corresponds to a single use of the channel. As
a consequence, the most general collective attack against a multi-way
scheme, even if incoherent between separate runs, may involve quantum
correlations between different channels. In general, an arbitrary collective
attack against a two-way scheme can be called a \emph{two-mode attack}. This
is the general scenario of Fig.~\ref{Pic1}(c) where the action of this
attack on a single round-trip of quantum communication is given by an
arbitrary map $\mathcal{E}^{2}$. On the one hand, such an attack is said to
be \emph{reducible} to a \emph{one-mode attack} if the map can be
symmetrically decomposed as $\mathcal{E}^{2}=\mathcal{E}\circ \mathcal{E}$
(i.e., the attack can be described by the scenario of Fig.~\ref{Pic1}(b)) On
the other hand, the two-mode attack is called $\emph{irreducible}$ if $%
\mathcal{E}^{2}\neq \mathcal{E}\circ \mathcal{E}$ (i.e., the attack of Fig.~%
\ref{Pic1}(c) cannot be described by Fig.~\ref{Pic1}(b)). The latter
situation includes all attacks where some kind of correlation is exploited
between the two paths, either if this correlation is classical (so that $%
\mathcal{E}^{2}=\mathcal{E}_{2}\circ \mathcal{E}_{1}$ with $\mathcal{E}%
_{1}\neq \mathcal{E}_{2}$) or truly quantum (so that $\mathcal{E}^{2}\neq
\mathcal{E}_{2}\circ \mathcal{E}_{1}$ for every $\mathcal{E}_{1}$ and $%
\mathcal{E}_{2}$).

In order to detect and handle an irreducible attack, the previous two-way
protocols, $Hom^{2}$ and $Het^{2}$, must be modified into hybrid forms that
we denote by $Hom^{1,2}$ and $Het^{1,2}$. In this hybrid formulation, Alice
randomly switches between a two-way scheme and the corresponding one-way
scheme, where she simply detects the incoming mode and sends a new one back
to Bob. We may describe this process by saying that Alice randomly closes
(ON) and opens (OFF) the quantum communication \emph{circuit} with Bob, the
effective switching sequence being communicated at the end of the protocol
(see Fig.~\ref{Pic1}(d)). By publicizing part of the exchanged data, Alice
and Bob can perform tomography of the quantum channels in both the ON and
OFF\ configurations. In particular, they can reconstruct the channel $%
\mathcal{E}^{2}$ affecting the two-way trip and the channels $\mathcal{E}%
_{1} $ and $\mathcal{E}_{2}$ affecting the forward and backward paths (see
Fig.~\ref{Pic1}(d)). Then, they can check the\emph{\ reducibility}\
conditions $\mathcal{E}_{1}=\mathcal{E}_{2}$ and $\mathcal{E}^{2}=\mathcal{E}%
_{2}\circ \mathcal{E}_{\alpha }\circ \mathcal{E}_{1}$, where $\mathcal{E}%
_{\alpha }(\rho )=\hat{D}(\alpha )\rho \hat{D}^{\dagger }(\alpha
)$ is Alice's publicized encoding map. If such conditions are
satisfied then the two-mode attack is reducible, i.e., Alice and
Bob have excluded every kind of quantum and classical correlation
between the two paths of the quantum communication (see Appendix
for an explicit description). In such a case, the honest users can
therefore exploit the superadditivity of the two-way quantum
communication. If the previous reducibility conditions are not
met, then the honest users can always exploit the instances of
one-way quantum communication. Notice that the verification of the
reducibility conditions is rather easy in the Gaussian case, where
the channels can be completely reconstructed by analyzing the
first and second statistical moments of the output states. Also
notice that the reducibility conditions exclude every kind of
quantum impersonation attack \cite{Dusek}, where Eve
short-circuits the channels of the two-way quantum communication.

In conclusion, the hybrid protocols constitute a safe implementation of
two-way protocols, at least in the presence of collective Gaussian attacks
(one-mode or two-mode). In the hybrid formulation, Alice and Bob can in fact
optimize their security on both one-way and two-way quantum communication.
The ON-OFF manipulation of the quantum communication can be interpreted as
if Alice had two orthogonal bases to choose from during the key distribution
process. In the presence of this randomization, Eve is not able to optimize
her Gaussian attack with respect to both kinds of quantum communication and
the trusted parties can always make the \emph{a posteriori} optimal choice.
As a natural development of these results, one can consider a situation
where Bob also performs a random and independent ON-OFF manipulation of the
quantum communication. Such a scheme naturally leads to instances of $n$-way
quantum communication (with $n>2$) whose security properties would be
interesting to inspect in future work. In general, our results pave the way
for future investigations in the domain of secure multiple quantum
communications, where quantum communication circuits can in principle grow
to higher and higher complexity.

\section{Acknowledgements}

The research of S. Pirandola was supported by a Marie Curie
Fellowship within the 6th European Community Framework Programme
(Contract No. MOIF-CT-2006-039703). S. Pirandola thanks CNISM for
hospitality at the University of Camerino and Gaetana Spedalieri
for her moral and logistic support. S. Lloyd was supported by the
W.M. Keck foundation center for extreme quantum information theory
(xQIT).

\appendix

\section{Appendix}

In this technical section, we first review some basic information about
Gaussian states. Then, we exhibit the general expressions for the secret-key
rates (in both direct and reverse reconciliation) when the various protocols
are subject to one-mode Gaussian attacks. In the following subsections, we
explicitly compute these secret-key rates for all the one-way and two-way
protocols. From these quantities we derive the security thresholds shown in
the paper. Finally, in the last subsection, we give the explicit description
of a general two-mode attack and we analyze the conditions for its
reducibility. This last analysis shows the security of the hybrid protocols
against collective Gaussian attacks.

\subsection{Basics of Gaussian states}

A bosonic system of $n$ modes can be described by a quadrature row-vector $%
\mathbf{\hat{Y}}:=(\hat{Q}_{1},\hat{P}_{1},\ldots ,\hat{Q}_{n},\hat{P}_{n})$
satisfying $[\hat{Y}_{l},\hat{Y}_{m}]=2i\Omega _{lm}$ ($1\leq l,m\leq 2n$),
where%
\begin{equation}
\mathbf{\Omega }:=\bigoplus\limits_{k=1}^{n}\left(
\begin{array}{cc}
0 & 1 \\
-1 & 0%
\end{array}%
\right)
\end{equation}%
defines a symplectic form. A bosonic state $\rho $ is called Gaussian if its
statistics is Gaussian \cite{EisertGauss,GaussianStates}. In such a case,
the state $\rho $ is fully characterized by its displacement $\langle
\mathbf{\hat{Y}}\rangle =\mathrm{Tr}(\mathbf{\hat{Y}}\rho )$ and correlation
matrix (CM) $\mathbf{V}$, whose generic element is defined by $%
V_{lm}:=\langle \hat{Y}_{l}\hat{Y}_{m}+\hat{Y}_{m}\hat{Y}_{l}\rangle
/2-\langle \hat{Y}_{l}\rangle \langle \hat{Y}_{m}\rangle $ with diagonal
terms $V_{ll}=\langle \hat{Y}_{l}^{2}\rangle -\langle \hat{Y}_{l}\rangle
^{2}:=V(\hat{Y}_{l})$ express the variances of the quadratures. According to
Williamson's theorem \cite{Williamson}, every CM $\mathbf{V}$ can be put in
diagonal form by means of a symplectic transformation, i.e., there exists a
matrix $\mathbf{S}$, satisfying $\mathbf{\mathbf{S}}^{T}\mathbf{\Omega S}=%
\mathbf{\Omega }$, such that $\mathbf{\mathbf{S}}^{T}\mathbf{VS}=\mathbf{%
\Delta (}\nu _{1},\nu _{1},\cdots ,\nu _{n},\nu _{n})$, where $\mathbf{%
\Delta }$ denotes a diagonal matrix. The set of real values $\boldsymbol{\nu
}:=\{\nu _{1},\cdots ,\nu _{n}\}$ is called \emph{symplectic eigenspectrum}
of the CM and provides compact ways to express fundamental properties of the
corresponding Gaussian state. In particular, the Von Neumann entropy $S(\rho
):=-\mathrm{Tr}\left( \rho \log \rho \right) $ of a Gaussian state $\rho $
can be expressed in terms of the symplectic eigenvalues by the formula \cite%
{Entropia}%
\begin{equation}
S(\rho )=\sum\limits_{k=1}^{n}g(\nu _{k})~,  \label{VN_Gauss}
\end{equation}%
where%
\begin{gather}
g(\nu ):=\tfrac{\nu +1}{2}\log \tfrac{\nu +1}{2}-\tfrac{\nu -1}{2}\log
\tfrac{\nu -1}{2}  \notag \\
\rightarrow \log \tfrac{e\nu }{2}+O(\nu ^{-1})\text{~~for }\nu \gg 1\text{~.}
\label{g_explicit}
\end{gather}%
Here, the information unit is the $\emph{bit}$ if $\log =\log _{2}$\ or the
\emph{nat} if $\log =\ln $.

An example of a Gaussian state is the two-mode squeezed vacuum state \cite%
{QObook} (or EPR source) whose CM takes the form%
\begin{equation}
\mathbf{V}_{EPR}(V)=\left(
\begin{array}{cc}
V\mathbf{I} & \sqrt{V^{2}-1}\mathbf{Z} \\
\sqrt{V^{2}-1}\mathbf{Z} & V\mathbf{I}%
\end{array}%
\right) ,  \label{EPR_source}
\end{equation}%
where $\mathbf{Z}:=\mathbf{\Delta }(1,-1)$ and $\mathbf{I}:=\mathbf{\Delta }%
(1,1)$. In Eq.~(\ref{EPR_source}) the variance $V$ fully characterizes the
EPR source \cite{QObook}. On the one hand, it quantifies the amount of
entanglement which is distributed between Alice and Bob, providing a
log-negativity \cite{LogNeg} equal to
\begin{eqnarray}
E_{\mathcal{N}} &=&\max \left\{ 0,-\tfrac{1}{2}\log (2V^{2}-1-2V\sqrt{V^{2}-1%
})\right\}  \notag \\
&\rightarrow &\log 2V+O(V^{-2})\text{~~for }V\gg 1\text{~.}
\end{eqnarray}%
On the other hand, it quantifies the amount of energy which is distributed
to the parties, since the reduced thermal states $\rho _{A}:=\mathrm{Tr}%
_{B}(\rho )$ and $\rho _{B}:=\mathrm{Tr}_{A}(\rho )$ have mean excitation
numbers equal to $(V-1)/2$.

\subsection{General expressions for the secret-key rates}

The various protocols differ for the number of paths ($1$ or $2$) and the
decoding method, which can be joint, disjoint, individual or collective. In
particular, when decoding is disjoint the relevant secret variable $X$ is $%
Q\in \mathbb{R}$ (or $P\in \mathbb{R}$, equivalently). When decoding is
joint, the relevant secret variable $X$ is $\{Q,P\}\in \mathbb{R}^{2}$.
Under the assumption of one-mode Gaussian attacks, the individual protocols (%
$Hom,Het,Hom^{2},Het^{2}$)\ have the following secret-key rates for DR ($%
\blacktriangleright $) and RR ($\blacktriangleleft $) \cite{DWrate,DWrate2}%
\begin{equation}
R^{\blacktriangleright }:=I(X_{A}:X_{B})-I(X_{A}:E)~,  \label{R_ind_DR}
\end{equation}%
\begin{equation}
R^{\blacktriangleleft }:=I(X_{A}:X_{B})-I(X_{B}:E)~.  \label{R_ind_RR}
\end{equation}%
In these formulae, $I(X_{A}:X_{B}):=H(X_{B})-H(X_{B}|X_{A})$ is the
classical mutual information between Alice and Bob's variables $X_{A}$ and $%
X_{B}$, with $H(X_{B})=(1/2)\log V(X_{B})$ and $H(X_{B}|X_{A})=(1/2)\log
V(X_{B}|X_{A})$ being the total and conditional Shannon entropies \cite%
{Shannon}. The term%
\begin{equation}
I(X_{K}:E):=H(E)-H(E|X_{K})  \label{Holevo_Bound}
\end{equation}%
is the Holevo information \cite{HInfo} between Eve ($E$) and the honest user
$K=A,B$ (i.e., Alice or Bob). Here, $H(E):=S(\rho _{E})$ is the Von Neumann
entropy of Eve's state $\rho _{E}$ and $H(E|X_{K})$ is the Von Neumann
entropy conditioned to the classical communication of $X_{K}$. For the
collective protocols ($\otimes Hom,\otimes Het,\otimes Hom^{2},\otimes
Het^{2}$) we have instead%
\begin{equation}
R^{\blacktriangleright }:=I(X_{A}:B)-I(X_{A}:E)~,  \label{R_coll_DR}
\end{equation}%
and%
\begin{equation}
R^{\blacktriangleleft }:=I(X_{A}:B)-I(B:E)~,  \label{R_coll_RR}
\end{equation}%
where $I(X_{A}:B)$, $I(X_{A}:E)$ are Holevo informations, and%
\begin{equation}
I(B:E):=H(B)+H(E)-H(B,E)  \label{Q_mutual_Info}
\end{equation}%
is the quantum mutual information between Bob and Eve. By setting $R=0$ in
the above Eqs.~(\ref{R_ind_DR}), (\ref{R_ind_RR}), (\ref{R_coll_DR}) and~(%
\ref{R_coll_RR}) one finds the security thresholds for the corresponding
protocols. Notice that the Holevo information of Eq.~(\ref{Holevo_Bound})
provides an upper bound to Eve's accessible information. In the case of
collective protocols, Alice and Bob are able to reach the Holevo bound $%
I(X_{A}:B)$ only asymptotically. This is possible if Alice communicates to
Bob the optimal collective measurement to be made compatible with the
generated sequence of signal states and the detected noise in the channel.
Such a measurement will be an asymptotic projection on the codewords of a
random quantum code as foreseen by the Holevo--Schumacher--Westmoreland
(HSW) theorem \cite{HSW,HSW2}. Though such a measurement is highly complex,
it is in principle possible and the study of the collective DR secret-key
rate of Eq.~(\ref{R_coll_DR}) does make sense (it is also connected to the
notion of private classical capacity of Ref.~\cite{Privacy}). On the other
hand, the quantum mutual information of Eq.~(\ref{Q_mutual_Info}) provides a
bound which is too large in general, preventing a comparison between the
collective protocols in RR.

\subsection{Secret-key rates of the one-way protocols}

In the one-way protocols, Alice encodes two independent Gaussian variables $%
Q_{A},P_{A}$ in the quadratures $\hat{Q}_{A},\hat{P}_{A}$ of a signal mode $%
A $, i.e., $\hat{Q}_{A}=Q_{A}+\hat{Q}_{A}|Q_{A}$ and $\hat{P}_{A}=P_{A}+\hat{%
P}_{A}|P_{A}$. Here, the quantum variables $\hat{Q}_{A},\hat{P}_{A}$ have a
global modulation $V$, given by the sum of the classical modulation $%
V(Q_{A})=V(P_{A})=V-1$ and the quantum shot-noise $V(\hat{Q}_{A}|Q_{A})=V(%
\hat{P}_{A}|P_{A})=1$. On the other hand, Eve has an EPR source $\mathbf{V}%
_{EPR}(W)$ which distributes entanglement between modes $E$ and $E^{\prime
\prime }$. The spy mode $E$ is then mixed with the signal mode $A$ via a
beam splitter of transmission $T$, and the output modes, $B$ and $E^{\prime
} $, are received by Eve and Bob, respectively (see Fig.~\ref{OneWayPic}).
Let us first consider the case of collective protocols ($\otimes Hom,\otimes
Het$), where Bob performs a coherent detection on all the collected modes $B$
in order to decode $X_{A}=Q_{A}$ (for $\otimes Hom$) or $X_{A}=\{Q_{A},P_{A}%
\}$ (for $\otimes Het$). For an arbitrary triplet $\{V,W,T\}$, the
quadratures of the output modes, $B$ and $E^{\prime }$, have variances%
\begin{gather}
V(\hat{Q}_{B})=V(\hat{P}_{B})=(1-T)W+TV:=b_{V}~,  \label{Bob_var_total} \\
V(\hat{Q}_{E^{\prime }})=V(\hat{P}_{E^{\prime }})=(1-T)V+TW:=e_{V}~,
\label{Eve_var_total}
\end{gather}%
and conditional variances%
\begin{gather}
V(\hat{Q}_{B}|Q_{A})=V(\hat{P}_{B}|P_{A})=(1-T)W+T=b_{1}~,
\label{Bob_var_cond} \\
V(\hat{Q}_{E^{\prime }}|Q_{A})=V(\hat{P}_{E^{\prime
}}|P_{A})=(1-T)+TW=e_{1}~.  \label{Eve_var_cond}
\end{gather}%
Globally, the CMs of the output states $\rho _{B}$ (of Bob), $\rho
_{E^{\prime }E^{\prime \prime }}:=\rho _{E}$ (of\ Eve) and $\rho _{E^{\prime
}E^{\prime \prime }B}:=\rho _{EB}$ (of Eve and Bob) are given by%
\begin{gather}
\mathbf{V}_{B}(V,V)=\mathbf{\Delta (}b_{V},b_{V})~, \\
\mathbf{V}_{E}(V,V)=%
\begin{pmatrix}
\mathbf{\Delta \lbrack }e_{V},e_{V}] & \varphi \mathbf{Z} \\
\varphi \mathbf{Z} & W\mathbf{I}%
\end{pmatrix}%
~,
\end{gather}%
and%
\begin{equation}
\mathbf{V}_{EB}=\left(
\begin{array}{cc}
\mathbf{V}_{E} & \mathbf{F} \\
\mathbf{F}^{T} & \mathbf{V}_{B}%
\end{array}%
\right) ~,~\mathbf{F:}=\binom{\mu \mathbf{I}}{\theta \mathbf{Z}}~,
\end{equation}%
where%
\begin{equation}
\varphi :=[T(W^{2}-1)]^{1/2}~,~\mu :=(W-V)[(1-T)T]^{1/2}~,
\end{equation}%
and%
\begin{equation}
\theta :=[(1-T)(W^{2}-1)]^{1/2}~.
\end{equation}%
The CMs of Bob ($B$) and Eve ($E$), conditioned to Alice's variable $X_{A}$,
are instead equal to%
\begin{equation}
\mathbf{V}_{K|Q_{A}}=\mathbf{V}_{K}(1,V)~,~\mathbf{V}_{K|Q_{A},P_{A}}=%
\mathbf{V}_{K}(1,1)~,
\end{equation}%
where $K=B,E.$ For $T\neq 0,1$ and $V\gg 1$, the symplectic spectra of all
the previous CMs are given by:%
\begin{eqnarray}
\boldsymbol{\nu }_{B} &\rightarrow &\{TV\}~, \\
\boldsymbol{\nu }_{B|Q_{A}} &\rightarrow &\{\sqrt{b_{1}TV}\}~, \\
\boldsymbol{\nu }_{B|Q_{A},P_{A}} &\rightarrow &\{b_{1}\}~, \\
\boldsymbol{\nu }_{E} &\rightarrow &\{(1-T)V,W\}~, \\
\boldsymbol{\nu }_{E|Q_{A}} &\rightarrow &\{\sqrt{e_{1}(1-T)V},\sqrt{%
Wb_{1}/e_{1}}\}~, \\
\boldsymbol{\nu }_{E|Q_{A},P_{A}} &\rightarrow &\{b_{1},1\}~, \\
\boldsymbol{\nu }_{BE} &\rightarrow &\{V,1,1\}~.
\end{eqnarray}%
By using Eqs.~(\ref{VN_Gauss}) and~(\ref{g_explicit}), we then compute all
the Von Neumann entropies to be used in the quantities $I(X_{A}:B),$ $%
I(X_{A}:E)$ and $I(B:E)$ of Eqs.~(\ref{R_coll_DR}) and~(\ref{R_coll_RR}).
After some algebra we get the following asymptotic rates for the one-way
collective protocols%
\begin{equation}
R^{\blacktriangleright }[\otimes Het]=\log \tfrac{T}{1-T}-g(W)~,
\end{equation}%
\begin{equation}
R^{\blacktriangleright }[\otimes Hom]=\tfrac{1}{2}\log \tfrac{Te_{1}}{%
(1-T)b_{1}}+g\left( \sqrt{\tfrac{Wb_{1}}{e_{1}}}\right) -g(W)~\text{,}
\end{equation}%
and%
\begin{equation}
R^{\blacktriangleleft }[\otimes Het]=\log \tfrac{1}{1-T}-g(W)-g(b_{1})~,
\end{equation}%
while $R^{\blacktriangleleft }[\otimes Hom]\rightarrow -\infty $, because of
the too large bound provided by $I(B:E)$ in this case.

Let us now consider the individual one-way protocols ($Hom,Het$). Bob's
output variable is $Q_{B}=\hat{Q}_{B}$ for $Hom$, and
\begin{equation}
\{Q_{B},P_{B}\}=2^{-1/2}\{\hat{Q}_{B}+\hat{Q}_{0},\hat{P}_{B}-\hat{P}_{0}\}
\end{equation}%
for $Het$ (with $\hat{Q}_{0},\hat{P}_{0}$ belonging to the vacuum). From
Eqs.~(\ref{Bob_var_total}) and~(\ref{Bob_var_cond}), we can calculate the
variances $V(X_{B})$ and $V(X_{B}|X_{A})$ that provide the non-computed term
$I(X_{A}:X_{B})$ in Eq.~(\ref{R_ind_DR}). Then, we get the following
asymptotic rates in DR%
\begin{equation}
R^{\blacktriangleright }[Hom]=R^{\blacktriangleright }[\otimes Hom]\text{~,}
\end{equation}%
and%
\begin{equation}
R^{\blacktriangleright }[Het]=\log \tfrac{2T}{e(1-T)(1+b_{1})}%
+g(b_{1})-g(W)~.
\end{equation}%
In order to derive the RR rates from Eq.~(\ref{R_ind_RR}) we must evaluate
\begin{equation}
I(X_{B}:E)=H(E)-H(E|X_{B})~,
\end{equation}%
where $H(E|X_{B})$ is computed from the spectrum $\boldsymbol{\nu }%
_{E|X_{B}} $ of the conditional CM $\mathbf{V}_{E|X_{B}}$. In RR, Eve's
quantum variables
\begin{equation}
\hat{Y}_{E}:=(\hat{Q}_{E^{\prime }},\hat{P}_{E^{\prime }},\hat{Q}_{E^{\prime
\prime }},\hat{P}_{E^{\prime \prime }})
\end{equation}%
must be conditioned to the Bob's classical variable $X_{B}.$ This is
equivalent to constructing, from $X_{B}$, the \emph{optimal }linear
estimators $\hat{Y}_{E}^{(X_{B})}$ of $\hat{Y}_{E}$, in such a way that the
residual conditional variables%
\begin{equation}
\hat{Y}_{E}|X_{B}:=\hat{Y}_{E}-\hat{Y}_{E}^{(X_{B})}
\end{equation}%
have minimal entropy $H(E|X_{B})$. For the $Hom$ protocol, Bob's variable $%
X_{B}=Q_{B}$ can be used to estimate the $\hat{Q}$ quadratures only. Then,
let Bob estimate $\hat{Y}_{E}$ by
\begin{equation}
\hat{Y}_{E}^{(Q_{B})}=(q^{\prime }Q_{B},0,q^{\prime \prime }Q_{B},0)~,
\end{equation}%
so that the conditional variables are given by
\begin{equation}
\hat{Y}_{E}|Q_{B}=(\hat{Q}_{E^{\prime }}-q^{\prime }Q_{B},\hat{P}_{E^{\prime
}},\hat{Q}_{E^{\prime \prime }}-q^{\prime \prime }Q_{B},\hat{P}_{E^{\prime
\prime }})~.
\end{equation}%
For $T\neq 0,1$ and $V\gg 1$, the optimal estimators are given by
\begin{equation}
q^{\prime }=-\sqrt{(1-T)/T}~,
\end{equation}%
and $q^{\prime \prime }=0$. The corresponding conditional spectrum%
\begin{equation}
\boldsymbol{\nu }_{E|Q_{B}}\rightarrow \{\sqrt{VW(1-T)/T},1\}
\end{equation}%
minimizes $H(E|Q_{B})$ and leads to the asymptotic rate%
\begin{equation}
R^{\blacktriangleleft }[Hom]=\tfrac{1}{2}\log \tfrac{W}{(1-T)b_{1}}-g(W)~.
\end{equation}%
For the $Het$ protocol, Bob's variable $X_{B}=\{Q_{B},P_{B}\}$ enables him
to estimate both the $\hat{Q}$ and $\hat{P}$ quadrature, by constructing the
$\hat{Y}_{E}$-linear estimator%
\begin{equation}
\hat{Y}_{E}^{(Q_{B},P_{B})}=(q^{\prime }Q_{B},p^{\prime }P_{B},q^{\prime
\prime }Q_{B},p^{\prime \prime }P_{B})~.
\end{equation}%
For $T\neq 0,1$ and $V\gg 1$, the optimal choice corresponds to
\begin{equation}
q^{\prime }=p^{\prime }=-\sqrt{2(1-T)/T}~,
\end{equation}%
and $q^{\prime \prime }=p^{\prime \prime }=0$, which gives%
\begin{equation}
\boldsymbol{\nu }_{E|Q_{B},P_{B}}\rightarrow \{(1-T+b_{1})/T,1\}~,
\end{equation}%
and leads to the asymptotic rate%
\begin{equation}
R^{\blacktriangleleft }[Het]=\log \tfrac{2T}{e(1-T)(1+b_{1})}+g\left( \tfrac{%
1-T+b_{1}}{T}\right) -g(W)~.
\end{equation}

\subsection{Secret-key rates of the two-way protocols}

In the EPR formulation of the two-way protocols (see Fig.~\ref{TwoWayPic}),
Bob assists the encoding via an EPR source $\mathbf{V}_{EPR}(V)$ that
distributes entanglement between mode $B_{1}$, which is kept, and mode $C_{1}
$, which is sent in the channel and undergoes the action of an entangling
cloner $(T,W):C_{1}\rightarrow A_{1}$. On the perturbed mode $A_{1}$, Alice
performs a Gaussian modulation by adding a stochastic amplitude $\alpha
=(Q_{A}+iP_{A})/2$ with $V(Q_{A})=V(P_{A})=\bar{V}$ and $\left\langle
Q_{A}P_{A}\right\rangle =0$. The modulated mode $A_{2}$ is then sent back
through the channel, where it undergoes the action of a second entangling
cloner $(T,W):A_{2}\rightarrow B_{2}$, where the output mode $B_{2}$ is
finally received by Bob. Let us first consider the collective two-way
protocols ($\otimes Hom^{2},\otimes Het^{2}$), where Bob performs an optimal
coherent measurement upon all the collected modes $B_{1},B_{2}$ in order to
decode $X_{A}=Q_{A}$ (for $\otimes Hom^{2}$) or $X_{A}=\{Q_{A},P_{A}\}$ (for
$\otimes Het^{2}$). For an arbitrary quadruplet $\{\bar{V},V,W,T\}$, the CMs
of the output states $\rho _{B_{1}B_{2}}:=\rho _{B}$ (of Bob) and $\rho
_{E_{1}^{\prime }E_{1}^{\prime \prime }E_{2}^{\prime }E_{2}^{\prime \prime
}}:=\rho _{E}$ (of\ Eve) are given by%
\begin{equation}
\mathbf{V}_{B}(\bar{V},\bar{V})=\left(
\begin{array}{cc}
V\mathbf{I} & T\sqrt{V^{2}-1}\mathbf{Z} \\
T\sqrt{V^{2}-1}\mathbf{Z} & \boldsymbol{\Lambda }_{B}(\bar{V},\bar{V})%
\end{array}%
\right) ~,
\end{equation}%
and%
\begin{equation}
\mathbf{V}_{E}(\bar{V},\bar{V})=\left(
\begin{array}{cccc}
e_{V}\mathbf{I} & \varphi \mathbf{Z} & \mu ^{\prime }\mathbf{I} & \mathbf{0}
\\
\varphi \mathbf{Z} & W\mathbf{I} & \theta ^{\prime }\mathbf{Z} & \mathbf{0}
\\
\mu ^{\prime }\mathbf{I} & \theta ^{\prime }\mathbf{Z} & \boldsymbol{\Lambda
}_{E}(\bar{V},\bar{V}) & \varphi \mathbf{Z} \\
\mathbf{0} & \mathbf{0} & \varphi \mathbf{Z} & W\mathbf{I}%
\end{array}%
\right) ~,
\end{equation}%
where%
\begin{equation}
\boldsymbol{\Lambda }_{B}(\bar{V},\bar{V}):=[T^{2}V+(1-T^{2})W]\mathbf{I}+T%
\mathbf{\Delta (}\bar{V},\bar{V})~,
\end{equation}%
and%
\begin{equation}
\boldsymbol{\Lambda }_{E}(\bar{V},\bar{V}):=\gamma \mathbf{I+}(1\mathbf{-}T)%
\mathbf{\Delta (}\bar{V},\bar{V})~,
\end{equation}%
with%
\begin{equation}
\mu ^{\prime }:=-\sqrt{1-T}\mu ~,~\theta ^{\prime }:=-\sqrt{1-T}\theta ~,
\end{equation}%
and%
\begin{equation}
\gamma :=T(1-T)V+(1-T)^{2}W+TW~.
\end{equation}%
The CMs of Bob ($B$) and Eve ($E$), conditioned to Alice's variable $X_{A}$,
are instead equal to
\begin{equation}
\mathbf{V}_{K|Q_{A}}=\mathbf{V}_{K}(0,\bar{V})~,~\mathbf{V}_{K|Q_{A},P_{A}}=%
\mathbf{V}_{K}(0,0)~,
\end{equation}%
for $K=B,E$. Let us consider identical resources between Alice and Bob,
i.e., $\bar{V}=V-1$. Then, for $T\neq 0,1$ and $V\gg 1$, all the symplectic
spectra are given by:%
\begin{eqnarray}
\boldsymbol{\nu }_{B} &\rightarrow &\{f_{1}V,f_{2}V\}~, \\
\boldsymbol{\nu }_{B|Q_{A}} &\rightarrow &\{\varsigma V,\varsigma ^{-1}\sqrt{%
T(1-T^{2})WV}\}~, \\
\boldsymbol{\nu }_{B|Q_{A},P_{A}} &\rightarrow &\{(1-T^{2})V,W\}~, \\
\boldsymbol{\nu }_{E} &\rightarrow &\{h_{1}V,h_{2}V,W,W\}~, \\
\boldsymbol{\nu }_{E|Q_{A}} &\rightarrow &\{\upsilon (1-T)V,\tfrac{\sqrt{%
(1-T^{2})WV}}{\upsilon },W,1\}~, \\
\boldsymbol{\nu }_{E|Q_{A},P_{A}} &\rightarrow &\{(1-T^{2})V,W,1,1\}~,
\end{eqnarray}%
where $f_{1}f_{2}=T$, $h_{1}h_{2}=(1-T)^{2}$ and
\begin{equation}
\varsigma :=[1+T^{2}(T^{2}+T-2)]^{1/2}~,~\upsilon :=[1+3T+T^{2}]^{1/2}~.
\end{equation}%
By means of Eqs.~(\ref{VN_Gauss}) and~(\ref{g_explicit}), we compute all the
Von Neumann entropies to be used in Eq.~(\ref{R_coll_DR}), and we get the
asymptotic rates%
\begin{equation}
R^{\blacktriangleright }[\otimes Hom^{2}]=\tfrac{1}{2}\log \tfrac{T}{%
(1-T)^{2}}-g(W)~,
\end{equation}%
and%
\begin{equation}
R^{\blacktriangleright }[\otimes Het^{2}]=2R^{\blacktriangleright }[\otimes
Hom^{2}]~.
\end{equation}%
Clearly these rates imply the same DR threshold $N^{\blacktriangleright
}=N^{\blacktriangleright }(T)$ for $\otimes Hom^{2}$ and $\otimes Het^{2}$,
as is shown in Fig.~\ref{DirectPic}. The derivation of $R^{%
\blacktriangleleft }[\otimes Het^{2}]$ and $R^{\blacktriangleleft }[\otimes
Hom^{2}]$ is here omitted because of the trivial negative divergence caused
by $I(B:E).$

Let us now consider the individual two-way protocols $Hom^{2}$ and $Het^{2}$
in DR. For the $Hom^{2}$ protocol, Bob decodes $Q_{A}$ by constructing the
output variable $Q_{B}:=Q_{B_{2}}-TQ_{B_{1}}$ from the measurements of $\hat{%
Q}_{B_{1}}$ and $\hat{Q}_{B_{2}}$. In fact, since $\hat{Q}%
_{B_{1}}\rightarrow \hat{Q}_{C_{1}}$ (for $V\gg 1$) and
\begin{equation}
\hat{Q}_{B_{2}}=\sqrt{T}Q_{A}+T\hat{Q}_{C_{1}}+\sqrt{1-T}(\sqrt{T}\hat{Q}%
_{E_{1}}+\hat{Q}_{E_{2}})~,
\end{equation}%
we asymptotically have
\begin{equation}
Q_{B}\rightarrow \sqrt{T}Q_{A}+\delta Q~,
\end{equation}%
with%
\begin{equation}
\delta Q:=\sqrt{1-T}(\sqrt{T}\hat{Q}_{E_{1}}+\hat{Q}_{E_{2}})~.
\end{equation}%
From $V(Q_{B})$ and $V(Q_{B}|Q_{A})$ we easily compute $I(Q_{A}:Q_{B})$ to
be used in Eq.~(\ref{R_ind_DR}). It is then easy to check that the
asymptotic DR rate satisfies%
\begin{equation}
R^{\blacktriangleright }[Hom^{2}]=R^{\blacktriangleright }[\otimes Hom^{2}]~.
\end{equation}%
For the $Het^{2}$ protocol, Bob measures
\begin{equation}
\left\{
\begin{array}{l}
\hat{q}_{-}=2^{-1/2}(\hat{Q}_{B_{1}}-\hat{Q}_{0})~, \\
\hat{p}_{+}=2^{-1/2}(\hat{P}_{B_{1}}+\hat{P}_{0})~,%
\end{array}%
\right.
\end{equation}%
from the first heterodyne on $B_{1}$, and%
\begin{equation}
\left\{
\begin{array}{l}
\hat{Q}_{-}=2^{-1/2}(\hat{Q}_{B_{2}}-\hat{Q}_{0^{\prime }})~, \\
\hat{P}_{+}=2^{-1/2}(\hat{P}_{B_{2}}+\hat{P}_{0^{\prime }})~,%
\end{array}%
\right.
\end{equation}%
from the second one upon $B_{2}$. Then, Bob decodes $\{Q_{A},P_{A}\}$\ via
the variables%
\begin{equation}
\left\{
\begin{array}{l}
Q_{B}:=Q_{-}-Tq_{-}~, \\
P_{B}:=P_{+}+Tp_{+}~.%
\end{array}%
\right.
\end{equation}%
In fact, for $T\neq 0,1$ and $V\gg 1$, we have%
\begin{equation}
Q_{B}\rightarrow \sqrt{T/2}Q_{A}+\delta Q^{\prime }~,~P_{B}\rightarrow \sqrt{%
T/2}P_{A}+\delta P~,
\end{equation}%
with%
\begin{equation}
\delta Q^{\prime }:=2^{-1/2}(\delta Q+T\hat{Q}_{0}-\hat{Q}_{0^{\prime }})~,
\end{equation}%
and%
\begin{equation}
\delta P:=2^{-1/2}[\sqrt{1-T}(\sqrt{T}\hat{P}_{E_{1}}+\hat{P}_{E_{2}})+T\hat{%
P}_{0}+\hat{P}_{0^{\prime }}]~.
\end{equation}%
From $V(X_{B})=V(Q_{B})V(P_{B})$ and $%
V(X_{B}|X_{A})=V(Q_{B}|Q_{A})V(P_{B}|P_{A})$ we then compute $I(X_{A}:X_{B})$
and the consequent asymptotic DR rate%
\begin{equation}
R^{\blacktriangleright }[Het^{2}]=\log \tfrac{2T(1+T)}{%
e(1-T)[1+T^{2}+(1-T^{2})W]}-g(W)~.
\end{equation}

Let us now consider $Hom^{2}$ and $Het^{2}$ in RR. In order to derive the
corresponding\ rates from Eq.~(\ref{R_ind_RR}), we must again compute $%
H(E|X_{B})$ from the spectrum of the conditional CM $\mathbf{V}_{E|X_{B}}$,
where Eve's quantum variables%
\begin{equation}
\hat{Y}_{E}:=(\hat{Q}_{E_{1}^{\prime }},\hat{P}_{E_{1}^{\prime }},\hat{Q}%
_{E_{1}^{\prime \prime }},\hat{P}_{E_{1}^{\prime \prime }},\hat{Q}%
_{E_{2}^{\prime }},\hat{P}_{E_{2}^{\prime }},\hat{Q}_{E_{2}^{\prime \prime
}},\hat{P}_{E_{2}^{\prime \prime }})
\end{equation}%
are conditioned to Bob's output variable $X_{B}$. Of course, this is again
equivalent to finding the optimal linear estimators $\hat{Y}_{E}^{(X_{B})}$
of $\hat{Y}_{E}$. For the $Hom^{2}$ protocol where $X_{B}=Q_{B}$, the linear
estimators of $\hat{Y}_{E}$\ take the form
\begin{equation}
\hat{Y}_{E}^{(Q_{B})}:=(q_{1}^{\prime }Q_{B},0,q_{1}^{\prime \prime
}Q_{B},0,q_{2}^{\prime }Q_{B},0,q_{2}^{\prime \prime }Q_{B},0)~.
\end{equation}%
For $T\neq 0,1$ and $V\gg 1$, the optimal ones are given by$\ q_{1}^{\prime
}=q_{1}^{\prime \prime }=q_{2}^{\prime \prime }=0$ and $q_{2}^{\prime }=-%
\sqrt{(1-T)/T}$. The corresponding conditional spectrum is given by
\begin{equation}
\boldsymbol{\nu }_{E|Q_{B}}\rightarrow \{m_{1}V,m_{2}\sqrt{V},W,1\}~,
\end{equation}%
with%
\begin{equation}
m_{1}m_{2}=[T^{-1}(1-T)^{3}(1+T^{3})W]^{1/2}~.
\end{equation}%
This leads to the asymptotic rate%
\begin{equation}
R^{\blacktriangleleft }[Hom^{2}]=\tfrac{1}{2}\log \tfrac{1-T+T^{2}}{(1-T)^{2}%
}-g(W)~.
\end{equation}%
For the $Het^{2}$ protocol, where $X_{B}=\{Q_{B},P_{B}\}$, we have%
\begin{gather}
\hat{Y}_{E}^{(Q_{B},P_{B})}:=  \notag \\
(q_{1}^{\prime }Q_{B},p_{1}^{\prime }P_{B},q_{1}^{\prime \prime
}Q_{B},p_{1}^{\prime \prime }P_{B},q_{2}^{\prime }Q_{B},p_{2}^{\prime
}P_{B},q_{2}^{\prime \prime }Q_{B},p_{2}^{\prime \prime }P_{B})~.
\end{gather}%
For $T\neq 0,1$ and $V\gg 1$, the optimal estimators are given by
\begin{equation}
q_{2}^{\prime }=p_{2}^{\prime }=-\sqrt{2(1-T)/T}~,
\end{equation}%
and
\begin{equation}
q_{1}^{\prime }=p_{1}^{\prime }=q_{1}^{\prime \prime }=p_{1}^{\prime \prime
}=q_{2}^{\prime \prime }=p_{2}^{\prime \prime }=0~.
\end{equation}%
The corresponding conditional spectrum is given by
\begin{equation*}
\boldsymbol{\nu }_{E|Q_{B},P_{B}}\rightarrow
\{n_{1},n_{2},n_{3},(1-T^{2})V\}~,
\end{equation*}%
where
\begin{equation}
n_{1}n_{2}n_{3}=\frac{[1+T^{3}+(1-T)(1+T^{2})W]W}{T(1+T)}~.
\end{equation}%
This spectrum leads to the final asymptotic rate%
\begin{gather}
R^{\blacktriangleleft }[Het^{2}]=\log \tfrac{2T(1+T)}{%
e(1-T)[1+T^{2}+(1-T^{2})W]}  \notag \\
+\sum\limits_{i=1}^{3}g(n_{i})-2g(W)~.
\end{gather}

\subsection{Structure of two-mode attacks}

\begin{figure}[tbph]
\begin{center}
\includegraphics[width=0.45\textwidth] {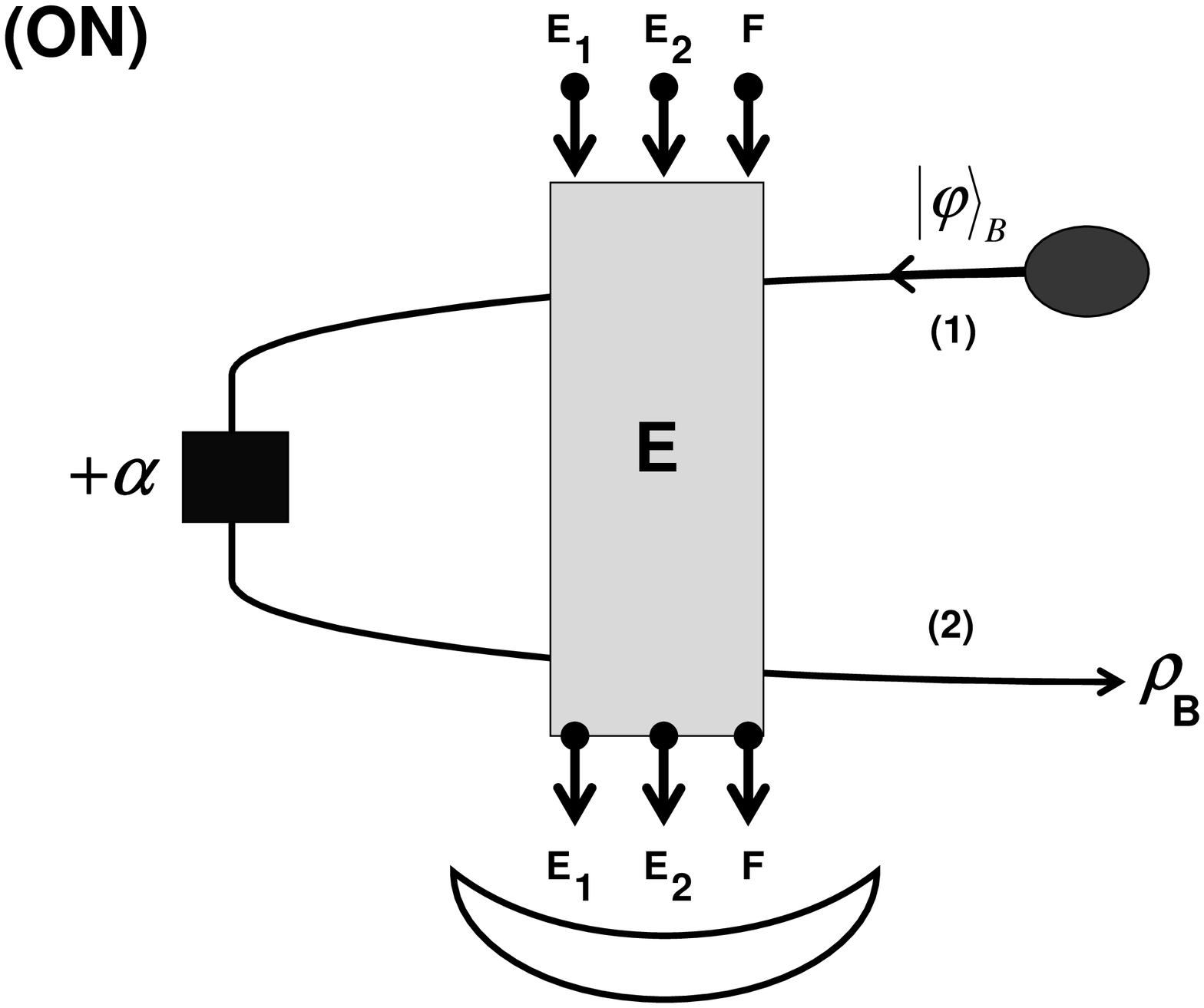}
\includegraphics[width=0.45\textwidth] {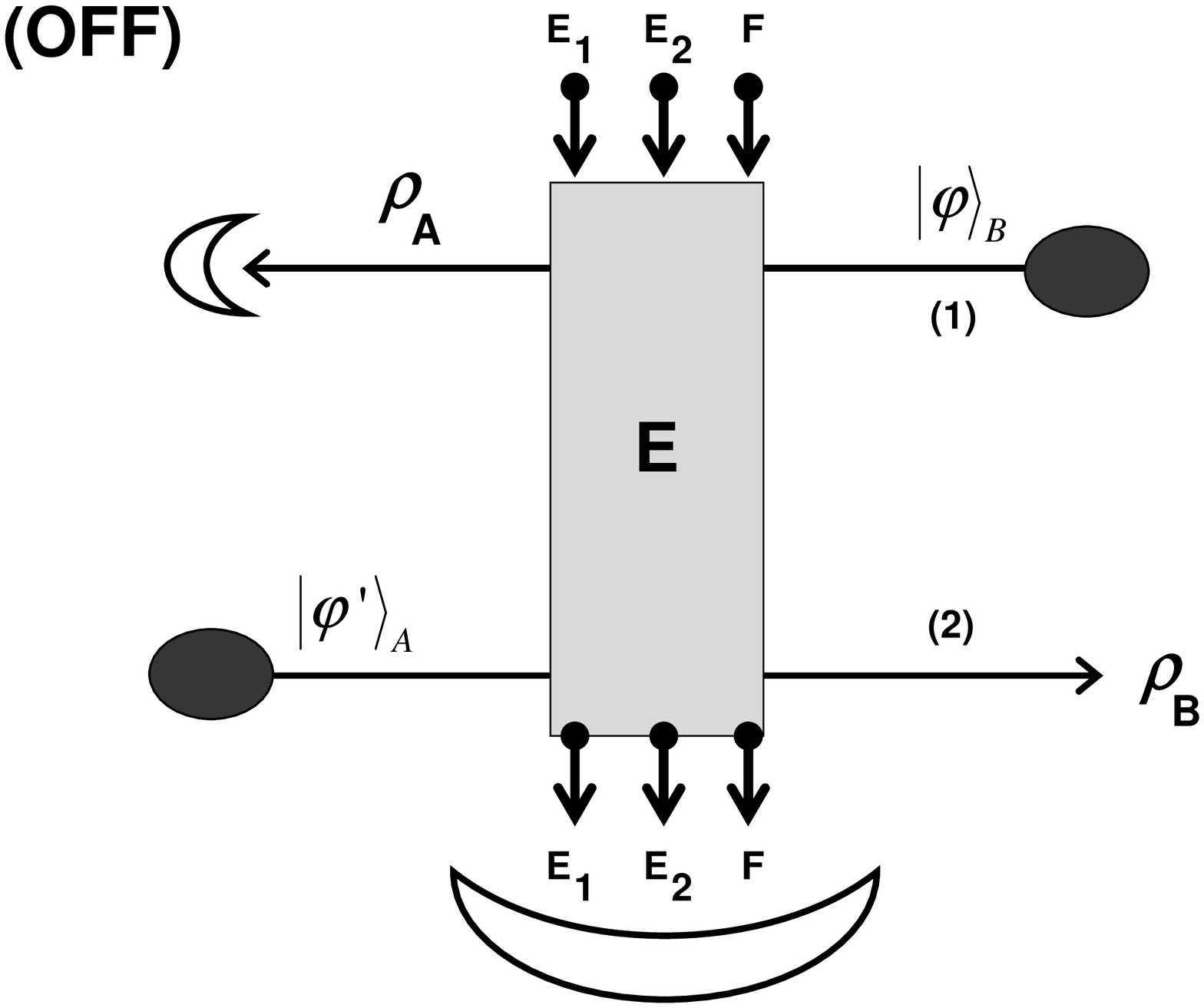}
\end{center}
\caption{General two-mode attack against a hybrid protocol (displayed in
both the ON and OFF configuration). The two paths of the quantum
communication interact with a supply of ancillas that can always be divided
into three blocks $E_{1},E_{2}$ and $F$.}
\label{Picirr}
\end{figure}
\begin{figure}[tbph]
\begin{center}
\includegraphics[width=0.45\textwidth] {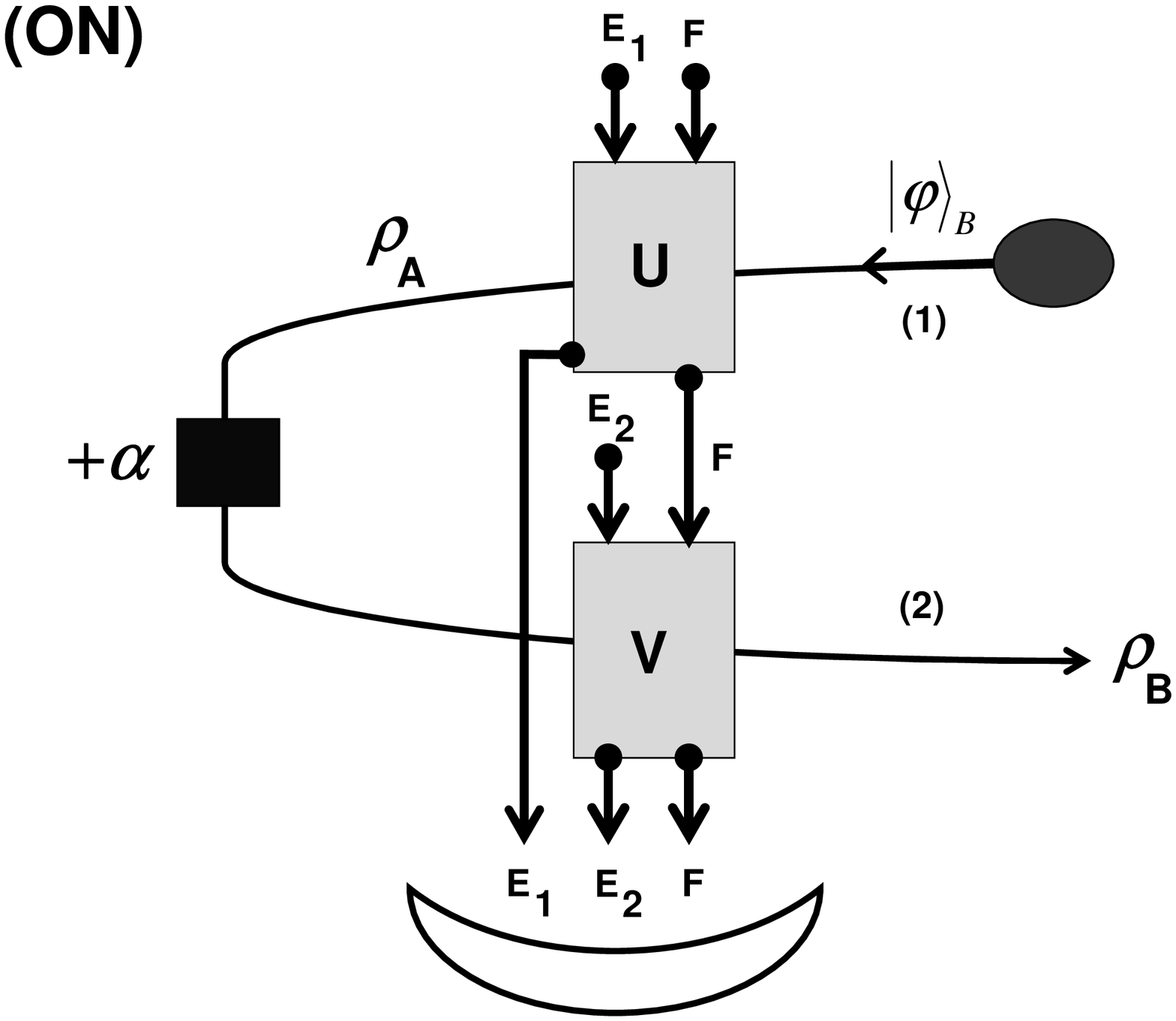}
\includegraphics[width=0.45\textwidth] {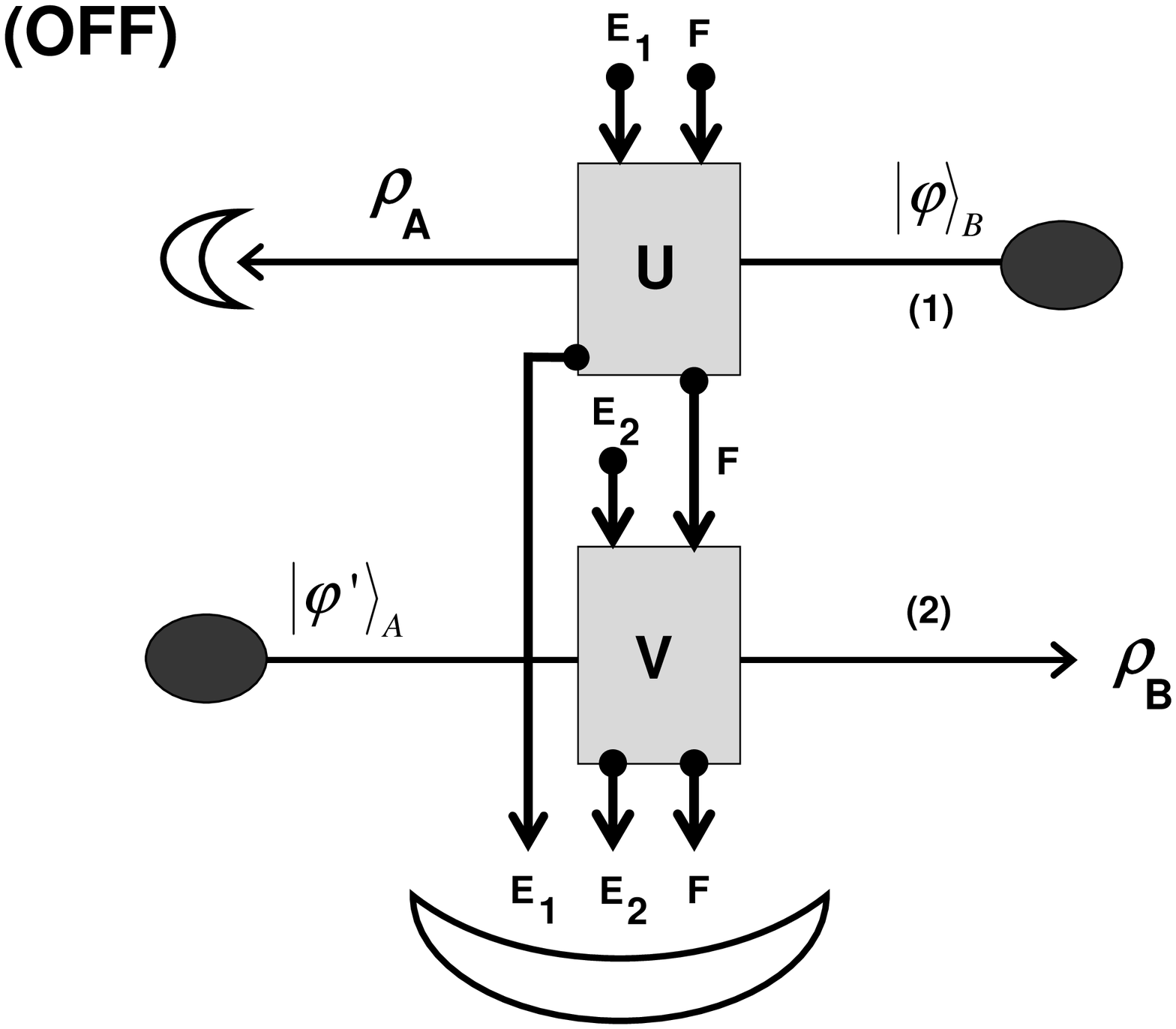}
\end{center}
\caption{Cascade form of the two-mode attack. In the first unitary $\hat{U}$%
, two blocks of ancillas $E_{1}$ and $F$ interact with the forward path
(labelled by 1). One output $E_{1}$ is sent to the final coherent detection
while the other one $F$ is taken as input for the second unitary $\hat{V}$.
Such a unitary makes the backward path (labelled by 2) interact with $F$
(coming from $\hat{U}$) and another block of fresh ancillas $E_{2}$. The
corresponding outputs of $F$ and $E_{2}$ are then sent to the final coherent
detection.}
\label{CollTWO}
\end{figure}

Let us describe the effects of a general two-mode attack when Alice and Bob
adopt the hybrid protocol. In the hybrid protocol, Bob sends a modulated
pure state $\left\vert \varphi \right\rangle _{B}$, which is coherent for $%
Het^{1,2}$\ and squeezed for $Hom^{1,2}$. Then, Alice modulates this state
by $\alpha $ in the ON configuration, while she detects $\left\vert \varphi
\right\rangle _{B}$ and re-sends a new $\left\vert \varphi ^{\prime
}\right\rangle _{A}$ in the OFF configuration. In general, in a two-mode
attack, Eve can use a countable set of ancillas which can always be
partitioned in three blocks $\mathbf{E}=\{E_{1},F,E_{2}\}$ (see Fig.~\ref%
{Picirr}). However, such an attack can always be reduced to the cascade form
of Fig.~\ref{CollTWO}. This is a trivial consequence of the logical
structure of the protocol, where the backward path (labelled by 2) is always
subsequent to the forward path (labelled by 1) and, therefore, a first
unitary interaction $\hat{U}$ can condition a second one $\hat{V}$, but the
contrary is not possible. In the first unitary $\hat{U}$, two blocks of
ancillas $E_{1}$ and $F$ interact with the forward path (1). One output $%
E_{1}$ is sent to the final coherent detection while the other one $F$ is
taken as input for the second unitary $\hat{V}$. Such a unitary makes the
backward path (2) interact with $F$ (coming from $\hat{U}$) and another
block of fresh ancillas $E_{2}$. The corresponding outputs of $F$ and $E_{2}$
are then sent to the final coherent detection. Note that such a description
contains all the possible quantum and/or classical correlations that Eve can
create between the forward and backward paths (both Gaussian and
non-Gaussian).

In the OFF\ configuration, the first channel $\mathcal{E}_{1}:\left\vert
\varphi \right\rangle _{B}\left\langle \varphi \right\vert \rightarrow \rho
_{A}$ is described by the Stinespring dilation \cite{Stines}%
\begin{gather}
\mathcal{E}_{1}(\left\vert \varphi \right\rangle _{B}\left\langle \varphi
\right\vert )=\mathrm{Tr}_{E_{1}F}\left[ \hat{U}_{BE_{1}F}\left( \left\vert
\varphi \right\rangle _{B}\left\langle \varphi \right\vert \right. \right.
\notag \\
\left. \left. \otimes \left\vert 0\right\rangle _{E_{1}}\left\langle
0\right\vert \otimes \left\vert 0\right\rangle _{F}\left\langle 0\right\vert
\right) \hat{U}_{BE_{1}F}^{\dagger }\right] ~,  \label{E_partial}
\end{gather}%
while the second channel $\mathcal{E}_{2}:\left\vert \varphi ^{\prime
}\right\rangle _{A}\left\langle \varphi ^{\prime }\right\vert \rightarrow
\rho _{B}$ can be expressed by the physical representation \cite{Holevo72}%
\begin{gather}
\mathcal{E}_{2}(\left\vert \varphi ^{\prime }\right\rangle _{A}\left\langle
\varphi ^{\prime }\right\vert )=\mathrm{Tr}_{E_{2}F}\left[ \hat{V}%
_{AE_{2}F}\left( \left\vert \varphi ^{\prime }\right\rangle _{A}\left\langle
\varphi ^{\prime }\right\vert \right. \right.  \notag \\
\left. \left. \otimes \rho _{F}\otimes \left\vert 0\right\rangle
_{E_{2}}\left\langle 0\right\vert \right) \hat{V}_{AE_{2}F}^{\dagger }\right]
~,  \label{second_map}
\end{gather}%
where
\begin{gather}
\rho _{F}=\mathrm{Tr}_{BE_{1}}\left[ \hat{U}_{BE_{1}F}\left( \left\vert
\varphi \right\rangle _{B}\left\langle \varphi \right\vert \right. \right.
\notag \\
\left. \left. \otimes \left\vert 0\right\rangle _{E_{1}}\left\langle
0\right\vert \otimes \left\vert 0\right\rangle _{F}\left\langle 0\right\vert
\right) \hat{U}_{BE_{1}F}^{\dagger }\right]
\end{gather}%
is the (generally mixed) state coming from the attack of the first channel.
In the ON configuration, the global map $\mathcal{E}^{2}:\left\vert \varphi
\right\rangle _{B}\left\langle \varphi \right\vert \rightarrow \rho _{B}$ is
equal to%
\begin{equation}
\mathcal{E}^{2}(\left\vert \varphi \right\rangle _{B}\left\langle \varphi
\right\vert )=\mathrm{Tr}_{\mathbf{E}}\left[ \hat{V}_{B\mathbf{E}}\left(
\left\vert \varphi \right\rangle _{B}\left\langle \varphi \right\vert
\otimes \left\vert 0\right\rangle _{\mathbf{E}}\left\langle 0\right\vert
\right) \hat{V}_{B\mathbf{E}}^{\dagger }\right] ~,  \label{Rev_SDtotal}
\end{equation}%
where $\left\vert 0\right\rangle _{\mathbf{E}}\left\langle 0\right\vert
:=\left\vert 0\right\rangle _{E_{1}}\left\langle 0\right\vert \otimes
\left\vert 0\right\rangle _{E_{2}}\left\langle 0\right\vert \otimes
\left\vert 0\right\rangle _{F}\left\langle 0\right\vert $ and%
\begin{equation}
\hat{V}_{B\mathbf{E}}:=\hat{V}_{BE_{2}F}\hat{D}_{B}(\alpha )\hat{U}%
_{BE_{1}F}~.  \label{Rev_SDunit}
\end{equation}%
Notice that the Stinespring dilation of Eq.~(\ref{Rev_SDtotal}) is \emph{%
unique} up to a local unitary transformation $\hat{U}_{\mathbf{E}}$ acting
on the output ancilla modes. In our description of the attack (see Fig.~\ref%
{CollTWO}) such a local unitary is included in the optimization of the final
coherent detection.

From Eq.~(\ref{Rev_SDunit}), it is clear that Eve's attack is void of \emph{%
quantum correlations} if%
\begin{equation}
\hat{V}_{BE_{2}F}=\hat{V}_{BE_{2}}\otimes \hat{V}_{F}~,  \label{Rev_first}
\end{equation}%
or
\begin{equation}
\hat{U}_{BE_{1}F}=\hat{U}_{BE_{1}}\otimes \hat{U}_{F}~.  \label{Rev_second}
\end{equation}%
In such a case in fact the two unitaries $\hat{U}$ and $\hat{V}$ are no
longer coupled by the $F$ ancillas. Let us assume one of the incoherence
conditions of Eq.~(\ref{Rev_first}) and~(\ref{Rev_second}). Then, we can
group the ancillas into two disjoint blocks $\mathbf{E}_{1}=\{E_{1},F\}$ and
$\mathbf{E}_{2}=\{E_{2}\}$ if Eq.~(\ref{Rev_first}) holds, or $\mathbf{E}%
_{1}=\{E_{1}\}$ and $\mathbf{E}_{2}=\{E_{2},F\}$ if Eq.~(\ref{Rev_second})
holds. In both cases the one-mode channels of Eqs.~(\ref{E_partial}) and~(%
\ref{second_map}) are expressed by the Stinespring dilations
\begin{equation}
\mathcal{E}_{1}(\left\vert \varphi \right\rangle _{B}\left\langle \varphi
\right\vert )=\mathrm{Tr}_{\mathbf{E}_{1}}\left[ \hat{U}_{B\mathbf{E}%
_{1}}\left( \left\vert \varphi \right\rangle _{B}\left\langle \varphi
\right\vert \otimes \left\vert 0\right\rangle _{\mathbf{E}_{1}}\left\langle
0\right\vert \right) \hat{U}_{B\mathbf{E}_{1}}^{\dagger }\right] ~,
\label{Rev_E1}
\end{equation}%
and%
\begin{gather}
\mathcal{E}_{2}(\left\vert \varphi ^{\prime }\right\rangle _{A}\left\langle
\varphi ^{\prime }\right\vert )=\mathrm{Tr}_{\mathbf{E}_{2}}\left[ \hat{V}_{A%
\mathbf{E}_{2}}\left( \left\vert \varphi ^{\prime }\right\rangle
_{A}\left\langle \varphi ^{\prime }\right\vert \right. \right.  \notag \\
\left. \left. \otimes \left\vert 0\right\rangle _{\mathbf{E}%
_{2}}\left\langle 0\right\vert \right) \hat{V}_{A\mathbf{E}_{2}}^{\dagger }%
\right] ~,  \label{Rev_E2}
\end{gather}%
while the two-mode channel $\mathcal{E}^{2}$ is expressed by Eq.~(\ref%
{Rev_SDtotal}) with%
\begin{equation}
\hat{V}_{B\mathbf{E}}:=\hat{V}_{B\mathbf{E}_{2}}\hat{D}_{B}(\alpha )\hat{U}%
_{B\mathbf{E}_{1}}~.  \label{Rev_Inco}
\end{equation}%
Now, one can easily check that Eq.~(\ref{Rev_Inco}) is equivalent to the
decomposability condition%
\begin{equation}
\mathcal{E}^{2}=\mathcal{E}_{2}\circ \mathcal{E}_{\alpha }\circ \mathcal{E}%
_{1}~.  \label{Rev_Check}
\end{equation}%
This can be easily verified by inserting Eqs.~(\ref{Rev_E1}) and~(\ref%
{Rev_E2}) into the right hand side of Eq.~(\ref{Rev_Check}) and resorting to
the uniqueness property of the Stinespring dilation.

Once the presence of quantum correlations between the paths has been
excluded, every residual classical correlation can be excluded by
symmetrizing the forward and backward channels, i.e., by setting $\mathcal{E}%
_{1}=\mathcal{E}_{2}$, which is equivalent to relating the two unitaries $%
\hat{U}$ and $\hat{V}$ by a partial isometry. In conclusion, the
verification of the condition $\mathcal{E}^{2}=\mathcal{E}\circ \mathcal{E}%
_{\alpha }\circ \mathcal{E}$ by Alice and Bob explicitly excludes every sort
of quantum/classical correlation between the two paths of the quantum
communication. Moreover, such a verification is relatively easy in case of
Gaussian attacks, since the corresponding Gaussian channels can be
completely reconstructed from the analysis of the first two statistical
moments.

\end{document}